\pdfoutput=1
\RequirePackage{ifpdf}
\ifpdf 
\documentclass[pdftex]{sigma}
\else
\documentclass{sigma}
\fi

\numberwithin{equation}{section}

\newtheorem{Theorem}{Theorem}[section]
\newtheorem{Lemma}[Theorem]{Lemma}
 { \theoremstyle{definition}
\newtheorem{Definition}[Theorem]{Definition} }

\usepackage{mathtools,mathrsfs,stmaryrd}
\usepackage{tensor}
\usepackage{upgreek}
\usepackage{bbm}

\newcommand{\mri}{\mathrm{i}}
\newcommand{\mre}{\mathrm{e}}
\newcommand{\mrd}{\mathrm{d}}

\newcommand{\mO}{\mathcal{O}}
\newcommand{\cC}{\mathcal{C}} 
\newcommand{\D}{\mathcal{D}} 
\newcommand{\cL}{\mathcal{L}}
\newcommand{\cH}{\mathcal{H}}
\newcommand{\dS}{\mathrm{d}\mathbb{S}}
\newcommand{\M}{\mathbb{M}}
\newcommand{\fC}{\mathfrak{C}}
\newcommand{\sA}{\mathscr{A}}
\newcommand{\fH}{\mathfrak{H}}
\newcommand{\frl}{\mathfrak{l}}
\newcommand{\fk}{\mathfrak{k}}
\newcommand{\fa}{\mathfrak{a}}
\newcommand{\fn}{\mathfrak{n}}
\newcommand{\fm}{\mathfrak{m}}

\newcommand{\fh}{\mathfrak{h}}
\newcommand{\U}{\mathcal{U}}
\newcommand{\cP}{\mathcal{P}}
\newcommand{\fp}{\mathfrak{p}}
\newcommand{\fb}{\mathfrak{b}}

\DeclareMathOperator{\ad}{ad}
\DeclareMathOperator{\grad}{grad}
\DeclareMathOperator{\Arg}{Arg}

\DeclareMathOperator{\sign}{sign}
\DeclareMathOperator{\sech}{sech}

\begin{document}
\allowdisplaybreaks

\newcommand{\arXivNumber}{1708.00538}

\renewcommand{\PaperNumber}{060}

\FirstPageHeading

\ShortArticleName{Wavepackets on de Sitter Spacetime}

\ArticleName{Wavepackets on de Sitter Spacetime}

\Author{Jo\~{a}o C.A. BARATA~$^\dag$ and Marcos BRUM~$^{\dag\ddag}$}

\AuthorNameForHeading{J.C.A.~Barata and M.~Brum}

\Address{$^\dag$~Instituto de F\'{\i}sica, Universidade de S\~{a}o Paulo, Rua do Mat\~{a}o 1371,\\
\hphantom{$^\dag$}~S\~{a}o Paulo, 05508-090, Brasil}
\EmailD{\href{mailto:jbarata@if.usp.br}{jbarata@if.usp.br}}

\Address{$^\ddag$~Departamento de Matem{\'a}tica, Universidade Federal do Rio de Janeiro,\\
\hphantom{$^\ddag$}~Campus Duque de Caxias, Rodovia Washington Luiz Km 104,5,\\
\hphantom{$^\ddag$}~Duque de Caxias, 25265-970, Brazil}
\EmailD{\href{mailto:marcos.brum@xerem.ufrj.br}{marcos.brum@xerem.ufrj.br}}

\ArticleDates{Received April 16, 2019, in final form August 12, 2019; Published online August 15, 2019}

\Abstract{We construct wavepackets on de Sitter spacetime, with masses consistently defined from the eigenvalues of an irreducible representation of a Casimir element in the universal enveloping algebra of the Lorentz algebra and analyse their asymptotic behaviour. Furthermore, we show that, in the limit as the de Sitter radius tends to infinity, the wave\-packets tend to the wavepackets of Minkowski spacetime and the plane waves arising after contraction have support sharply located on the mass shell.}


\Classification{81T20}

\section{Introduction}

The particle concept on a curved spacetime is ambiguous. This is intrinsically related to the absence of a preferred Hilbert space of states and also presents itself in the absence of an $S$-matrix. Some consequences of this ambiguity are illustrated in the Hawking and Unruh effects~\cite{Hawking74,Unruh76}.

We intend to take the first step in the formulation of a particle state (and of multiparticle asymptotic states in the sense of the Haag--Ruelle theory \cite{Dybalski17,DybalskiGerard14b,Haag58,Hepp65,Hepp_Brandeis65,Ruelle62}), namely, we want to construct and analyse the asymptotic behaviour of wavepackets on de Sitter spacetime. The first problem that we have met is the concept of mass, and it will be treated following~\cite{Bargmann47,Wigner39}. On Minkowski spacetime, the possible values of the mass of a field are the square roots of the elements of the spectrum of the representation of the Casimir element in the universal enveloping algebra of the Poincar\'{e} algebra which is physically interpreted as the square of the four-momentum, on the Hilbert space generated by smooth square integrable functions on Minkowski spacetime with invariant measure. There is also a Casimir element in the universal enveloping algebra of the Lorentz algebra which can be represented on a similar Hilbert space, and the elements in its spectrum can be related to the mass. The massive solutions of the wave equation on de Sitter spacetime have been analysed by several authors in different approaches \cite{Akhmedov14,BrosGazeauMoschella94,BrosMoschella96,BrosMoschella04, Dobrevetalli77,Garidi03,Harish-Chandra_I-58,Harish-Chandra_II-58,JMP_I-06, JMP_II-06,LimNiedRacz66a,LimNiedRacz67,Molchanov66,LimNiedRacz66b, Strichartz73,Takahashi63,Thieleker74}. The interpretation of these solutions as plane waves and the Fourier transform in the space of square integrable functions on de Sitter spacetime was presented in some of these references. However, wavepackets on de Sitter spacetime have not yet been defined in the literature. We are going to define wavepackets on de Sitter spacetime and prove some important properties.

Another desirable feature of any physical theory formulated on a curved spacetime is that it has a~sensible flat limit. More precisely, in the limit as the curvature of spacetime tends to zero, one must recover the corresponding physical theory formulated on Minkowski spacetime. In the context of Lie group theory, such a limit can be obtained under the technique of group contraction \mbox{\cite{DooleyRice85,Hannabuss71,InonuWigner53,MickNied72,Mizony84,Primet83}}. It is well known that, in this limit, the Lorentz group contracts towards the Poincar\'{e} group (of the same dimension). We will explore this fact to prove that the limit of a wavepacket on de Sitter spacetime is a wavepacket on Minkowski spacetime. The flat limit of the plane waves has been analysed by \cite{GarHugRen03,GazNov08}. The authors of \cite{BrosEpsteinGaudinMoschella10,BrosEpsteinMoschella98, BrosEpsteinMoschella08,BrosEpsteinMoschella10,MarolfMorrisonSrednicki13} analysed specific types of interactions by the use of perturbation methods.

The plan of the paper is as follows: in Section~\ref{sec_def} we present the definitions which will be used throughout the text; in Section~\ref{sec_repr-alg} we define the wave equation invariant under the group of isometries; in Section~\ref{sec_sols} we present the plane waves and the Fourier transform on de Sitter spacetime; in Section~\ref{sec_contract} we present our first important result, that the Casimir operator of the Lorentz algebra, whose eigenvalue is related to the mass of the plane wave on de Sitter spacetime, contracts towards the Casimir operator of the Poincar\'{e} algebra whose eigenvalue is related to the mass of the plane wave on Minkowski spacetime. This is a key result in the comparison between the wavepackets on de Sitter spacetime and the ones on Minkowski spacetime. Finally, in Section~\ref{sec_wpacket}, we construct the wavepackets on de Sitter spacetime, analyse their asymptotic behaviour and prove that, in the flat limit, they coincide with the usual wavepackets of the Minkowski spacetime. Moreover, the plane waves, after contraction, have support in momentum space sharply located on the mass shell. In Section~\ref{sec_concl}, we present our conclusions. The main results of the paper are the Theorems~\ref{wpack-fast} and~\ref{wpack-flat}:
\begin{enumerate}\itemsep=0pt
\item The wavepackets on de Sitter spacetime are functions of fast decrease.
\item \looseness=-1 The limit of a wavepacket on $n$-dimensional de Sitter spacetime, as the curvature radius of the spacetime tends to zero, is a wavepacket on Minkowski spacetime, analytic in the whole spacetime and with mass determined by the mass of its precedent wavepacket on~$\dS$.
\end{enumerate}

\section{Definitions}\label{sec_def}

The de Sitter spacetime ($\dS$) of dimension $n$ may be seen as a hyperboloid embedded in Minkowski spacetime $\M_{n+1}$ of dimension $n+1$ \cite{Schrodinger56}. Choosing a coordinate system which assigns to a point $p$ of $\M_{n+1}$ the point $\left(x_0(p), \ldots ,x_{n}(p)\right)$, the coordinates of any point in $\dS$ satisfy
\begin{gather}
x\cdot x \coloneqq -x_{0}^{2}+x_{1}^{2}+\dots +x_{n}^{2}=R^{2} ,\label{dS}
\end{gather}
where $R$ is the curvature of $\dS$ (also called the de Sitter radius). We will call this coordinate system ``ambient coordinate system''.

It is convenient to treat a particular point as the origin of the spacetime. Thus we define, without loss of generality, the point $\vartheta\in\dS$ with coordinates $(0,0,\ldots,0,R)$, to be the {\it origin of the de Sitter spacetime}. The origin of $\M_{n+1}$ is the point $o\in\M_{n+1}$ with coordinates $(0,0,\ldots,0,0)$. Moreover, the {\it null cone} $\fC$ in $\M_{n+1}$ is the locus of points which are connected to $o\in\M_{n+1}$ through null curves. It is important to remark that $\fC$ tangentiates the hyperboloid at infinity. The subset $\fC^+ \subset \fC$, called the {\it future null cone}, is the locus of points which are connected to $o\in\M_{n+1}$ through null curves whose tangent vectors are future directed.

The group of isometries of $\dS$ is the identity component of the Lorentz group $L\coloneqq\mathrm{SO}_0(1,n)$, a connected semi-simple Lie group, and the corresponding Lie algebra $\frl=\mathfrak{so}(1,n)$ is the Lorentz algebra. Its elements act on $\dS$ as rotations and hyperbolic rotations of the points. It is implemented by the map{\samepage
\begin{gather*}
\kappa\colon \ L \times \dS \ni (g,x) \mapsto xg \in \dS .
\end{gather*}
The $\dS$ is diffeomorphic to the homogeneous space $L/{\rm SO}_0(1,n-1)$.}

The corresponding representation $\Pi$ of the elements of $L$ as operators on the Hilbert space of smooth complex-valued square integrable functions $f\in\cL^2(\dS,\mrd\Sigma)$, where $\mrd\Sigma$ is the volume measure on $\dS$ (which is also invariant under the action of~$L$), is given by
\begin{gather}
(\Pi(g)f)(x)=\mathscr{D}_l(g)f(xg) ,\label{group_repr}
\end{gather}
where $\mathscr{D}_l$, in general, is a matrix and $l$ indexes the representation of $L$. In the present case, $\mathscr{D}_l$~is a scalar and~$l$ represents only the mass of~$f$ (to be defined below). On the other hand, the group of isometries of $\M_{n+1}$ is the Poincar\'{e} group $P_{n+1}\coloneqq\mathrm{SO}_0(1,n)\ltimes\mathbb{R}^{n+1}$. Its elements act on the points of $\M_{n+1}$ as rotations, hyperbolic rotations (``boosts''), and translations.

Another space of functions that will appear below is the space $\D(X)$ of smooth compactly supported functions on some space $X$ (to be precisely specified in each case).

Let now $\frl = \fk \oplus \fa \oplus \fn$ be an Iwasawa decomposition of the Lorentz algebra $\frl$, where $\fa$ is the maximal abelian subalgebra of $\frl$, $\fn$ the nilpotent subalgebra normalized by $\fa$ (i.e., for every $n \in \fn$ and $a \in \fa$, $[n,a] \in \fn$), $\fk$ the subalgebra on which the Cartan involution acts as the identity operator and $\fm \subset \fk$ the centralizer of $\fa$ in $\fk$ (i.e., for every $m \in \fm$ and $a \in \fa$, $[m,a]=0$). We remark that $\fm$ also normalizes $\fn$. $\mathscr{F}$ is the space of real functionals on $\fa$ and $\mathscr{F}_+ \subset \mathscr{F}$ the subset of positive real functionals. $\fn$ is identified as the disjoint union of positive root spaces $\fn=\bigoplus\limits_{\alpha\in\mathscr{F}_+}\frl_\alpha$. Furthermore, define $\rho\coloneqq(1/2)\sum\limits_{\alpha\in\mathscr{F}_+}\mathrm{m}(\alpha)\alpha$, where $\mathrm{m}(\alpha)$ is the dimension of the root space $\frl_\alpha$. Moreover, given two roots $\alpha$ and $\alpha'$, $(\alpha,\alpha')\coloneqq \beta(h_\alpha,h_{\alpha'})$, where $h_\alpha, h_{\alpha'} \in \fh$, $\fh$ is a toral Cartan subalgebra of~$\frl$~\cite{HilgertNeeb12} and $\beta(\cdot,\cdot)$ is the Cartan--Killing form of $\frl$. On $L$, this decomposition gives rise to the factorization $L=KAN=NAK$, where $K\coloneqq\langle\exp_{L}\fk\rangle$ is the compact group generated by $\exp_{L}\fk$, $A\coloneqq\exp_{L}\fa$ and $N\coloneqq\exp_{L}\fn$. $A$ and $N$ are simply connected closed subgroups of $L$ and $M\coloneqq\langle\exp_{L}\fm\rangle$.

A {\it horosphere} on $\dS$ is an orbit of the subgroup $N$. The action of $N$ on $T^\ast_o \M_{n+1}$ leaves invariant one generator of $\fC^+$. This generator is called the {\it normal to the horosphere}. Moreover, the generators of $\fC^+$ can be obtained one from the other by a rotation. We denote the set of all generators by $\sA$ and call this set the {\it absolute}.

There is an analogous definition of horosphere as a subset of the homogeneous space $L/{\rm SO}_0(1,\allowbreak n-1)$ (see \cite{Helgason68,Helgason84} and the discussion at the end of Section~9.2.1 in~\cite{Warner-II})\footnote{This definition can be extended to the quotient of any semisimple Lie group by one of its closed subgroups.}. A horosphere is an orbit of a subgroup of $L$ conjugate to~$N$. Since~$M$ normalizes~$N$ (i.e., for every $g \in M$, $gN=Ng$), this subgroup is $MN$. Denoting $\Xi$ the set of all horospheres in $L/{\rm SO}_0(1,n-1)$, $\Xi\cong L/MN$. In addition, the horospheres are closed submanifolds of $L/{\rm SO}_0(1,n-1)$. The origin of $L/{\rm SO}_0(1,n-1)$ is defined to be the left coset ${\rm eSO}_0(1,n-1) \equiv {\rm SO}_0(1,n-1)$, where $e$ is the identity element of $L$. Let $o\in {\rm SO}_0(1,n-1)$, $\xi_o=N.o$ is a horosphere passing through $o$. Any horosphere in $L/{\rm SO}_0(1,n-1)$ can be written as $kh.\xi_o \eqqcolon \xi_{kh}$, where $kM\in {\rm SO}_0(1,n-1)/M$ and $h\in A$ are unique. Conversely, $\forall\, kM\in {\rm SO}_0(1,n-1)/M$ and $\forall\, h\in A$, $kh.\xi_o$ is a horosphere in $L/{\rm SO}_0(1,n-1)$, hence $L$ permutes the horospheres transitively. $kM$ is called the {\it normal} to the horosphere $\xi_{kh}$ and $h$ is the {\it complex distance} from the origin $o$ to $\xi_{kh}$.

$\fa$ is actually one-dimensional, its generator is the generator of a hyperbolic rotation and $A$ leaves invariant one plane in $\dS$. Its action is given by the matrix ($\text{a}\in A$)
\begin{gather*}
\text{a}(\tau) \coloneqq \begin{pmatrix}
\cosh(\tau/R) & 0 & \ldots & 0 & \sinh(\tau/R) \\
0 & & \mathbbm{1}_{n-1} & & 0 \\
\sinh(\tau/R) & 0 & \ldots & 0 & \cosh(\tau/R)
\end{pmatrix} , \qquad \tau\in\mathbb{R}.
\end{gather*}
$\fn$ is an abelian subalgebra whose generators are the generators of the horospheric translations that leave invariant one of the null vectors $\xi$ in the tangent space of the plane left invariant by~$A$~\cite{GelGraVi66,VilKlim-II}. In the coordinate system~\eqref{dS}, $\xi$ has components $ (1,0,\ldots,0,1 )$. For $\vec{y}\in\mathbb{R}^{n-1}$ and $\text{n}\in N$,
\begin{gather*}
\text{n}\left(\vec{y}\right)=\begin{pmatrix}
\displaystyle 1+\frac{1}{2}\frac{|y|^{2}}{R^{2}} & \displaystyle \frac{1}{R}\vec{y} & \displaystyle\frac{1}{2}\frac{|y|^{2}}{R^{2}} \vspace{1mm}\\
\displaystyle\frac{1}{R}\vec{y}^{T} & \mathbbm{1}_{n-1} & \displaystyle\frac{1}{R}\vec{y}^{T}\vspace{1mm} \\
\displaystyle-\frac{1}{2}\frac{|y|^{2}}{R^{2}} &\displaystyle -\frac{1}{R}\vec{y} & \displaystyle1-\frac{1}{2}\frac{|y|^{2}}{R^{2}}
\end{pmatrix} .
\end{gather*}
Clearly, $\text{a}^{-1}(\tau)=\text{a}(-\tau)$ and $\text{n}^{-1}(\vec{y})=\text{n}(-\vec{y})$. Moreover, $\dim (\fn)=n-1$ and $\forall\, n\in\fn$, $\ad\fa(n)=n \therefore \alpha(\fa)=1$ and $\mathrm{m}(\alpha)=(n-1)/2$. $\fk=\mathfrak{so}(n)$ and $\fm\cong\mathfrak{so}(n-1)$, and the action of~$M$ leaves invariant each point of the plane left invariant by~$A$. $K$ and $M$ are, respectively, composed of matrices of the form
\begin{gather*}
K \coloneqq \begin{pmatrix}
1 & 0 \\
0 & \mathrm{SO}(n)
\end{pmatrix} , \qquad \textrm{and} \qquad M \coloneqq \begin{pmatrix}
1 & 0 & 0 \\
0 & \mathrm{SO}(n-1) & 0 \\
0 & 0 & 1
\end{pmatrix} .
\end{gather*}

In the following, we will always consider the indices $i$ and $j$ running through the set $\{1,\ldots,\allowbreak n-1\}$. We will also write $\text{n}_i \coloneqq \text{n}(y_{i})$, $m_{ij}\in M$ denoting rotation in the plane $x_i-x_j$ and $k_{in}\in K$ denoting rotation in the plane $x_i-x_n$. Moreover, almost every point of $\dS$ can be reached from the origin $\vartheta$ by the composition of a~hyperbolic rotation on the plane $x_{0}-x_{n}$, a~horospheric translation and a reflection:
\begin{align}
x(\tau,\vec{y})&=\vartheta \cdot \text{a}(\tau)\text{n}(\vec{y})\varepsilon_x \nonumber \\
&=R\left(\sinh(\tau/R)-\frac{1}{2}\frac{|y|^{2}}{R^{2}}\mre^{-\tau/R},-\frac{1}{R}\vec{y}\mre^{-\tau/R},\cosh(\tau/R)-\frac{1}{2}\frac{|y|^{2}}{R^{2}}\mre^{-\tau/R}\right)\varepsilon_x ,\label{horo_coord}
\end{align}
where $\varepsilon_x=\pm 1$. We remark that only points of the form $x_0+x_n=0$ are not covered by these charts, but these points form a set of measure zero in $\dS$. This coordinate system will be called ``horospheric coordinate system''. Another coordinate system that is going to be used later is the following parametrization of $\dS$:
\begin{align}
&x_0= R\sinh\beta , \nonumber \\
&x_1= R\cosh\beta\sin\varphi_1\cdots\sin\varphi_{n-2}\sin\phi, \nonumber \\
& x_2= R\cosh\beta\sin\varphi_1\cdots\sin\varphi_{n-2}\cos\phi , \nonumber \\
&\quad \vdotswithin{=} \notag \nonumber \\
& x_{n-1}= R\cosh\beta\sin\varphi_1\cos\varphi_{2} , \nonumber \\
& x_n= R\cosh\beta\cos\varphi_1 ,\label{hyper_coord}
\end{align}
where $\beta \in (-\infty,\infty)$, $\varphi_i \in [0,\pi]$ for $i \in \{1,\ldots,n-2\}$ and $\phi \in [0,2\pi)$. This parametrization will be called ``hyperbolic coordinate system''.

\section{Representations of the Lorentz algebra}\label{sec_repr-alg}

The action of $L$ on the points of $\dS$ induces, for every point $p\in\dS$, a homomorphism between~$\frl$ and the Lie algebra of vectors $v\in T_p\dS$~\cite{HilgertNeeb12}. If we choose the coordinate system described in~\eqref{dS} and denote the generators of $\frl$ by $m_{i0}$ (the hyperbolic rotations in the planes 0-$i$) and~$m_{ij}$ (the rotations in the planes $i$-$j$), with $i,j\in\{1,\ldots,n\}$, and consider a complex representation of the elements of $\frl$ as operators on the Hilbert space $\cL^2(\dS,\mrd\Sigma)$, these elements are represented as
\begin{gather}
\tensor{m}{_{i0}}=\mri\left(x_{i}\frac{\partial}{\partial x_0}+x_{0}\frac{\partial}{\partial x_i}\right) \qquad \mathrm{and} \qquad \tensor{m}{_{ij}}=\mri\left(x_{i}\frac{\partial}{\partial x_j}-x_{j}\frac{\partial}{\partial x_i}\right) . \label{m_i0-ij}
\end{gather}
However, it must be taken into account that the coordinate $x_n$ is constrained by \eqref{dS}, hence $\partial/\partial x_n$ can be written in terms of the derivatives with respect to the other coordinates. The commutation relations satisfied by the elements $m_{i0}$ and $m_{ij}$ are: let $\eta_{ab}=\mathrm{diag}\left[-1,1,\ldots, 1\right]$ be the metric tensor on $\M_{n+1}$,
\begin{gather*}
\left[m_{ab},m_{uv}\right]=\mri\left(\eta_{av}m_{bu}+\eta_{bu}m_{av}-\eta_{au}m_{bv}-\eta_{bv}m_{au}\right)={\text{C}_{abuv}}^{rs}m_{rs} ,
\end{gather*}
where ${\text{C}_{abuv}}^{rs}$ are the structure constants\footnote{The structure constants are displayed with twice the usual number of indices because each element of the Lie algebra is represented by a pair of indices.} of $\frl$.

Moreover, let $a \in \fa$ and $n_i \in \fn$ correspond to $\text{n}_i \in N$. These elements are written, in terms of the generators of $\frl$ presented above, as
\begin{gather}
a = m_{n0} \qquad \text{and} \qquad n_{i} = m_{i0}+m_{in} .\label{m-n}
\end{gather}

One Casimir element in the universal enveloping algebra $\U$ of $\frl$ is
\begin{gather*}
C^{2}\coloneqq j^{2}-m^{2} , \qquad \textrm{where} \qquad m^{2} \coloneqq \sum_{i=1}^{n}{m_{i0}}^{2} \qquad \mathrm{and} \qquad j^{2} \coloneqq \sum_{i<j}{m_{ij}}^{2} .
\end{gather*}
Moreover,
\begin{gather*}
\big[m^{2},j^{2}\big]=0 .
\end{gather*}
The representation described above is decomposed into a direct sum of irreducible representations, each labeled by one element of the spectrum of the operator ${\bf C}^2$, the representation of the Casimir element $C^{2}$ on $\cL^2(\dS,\mrd\Sigma)$. Let $-\mu^ {2}R^{2}$ be such an element.

The d'Alembert operator on $\cL^2(\dS,\mrd\Sigma)$ is related to the operator ${\bf C}^2$ by \cite{Dixmier61}
\begin{gather}
\Box_{\dS}=-\frac{1}{R^{2}}{\bf C}^{2}=\frac{1}{R^{2}}\left({\bf m}^{2}-{\bf j}^{2}\right) .\label{dalembert}
\end{gather}
Therefore a massive solution of the wave equation will be an eigenfunction of $\Box_{\dS}$ with eigenvalue~$\mu^ {2}$. The wave equation assumes the form
\begin{gather}
\big(\Box_{\dS}-\mu^ {2}\big)\psi = 0 .\label{wave_eq}
\end{gather}

\section{Solutions of the wave equation}\label{sec_sols}

One can verify, by direct inspection, that
\begin{gather}
\psi_{\xi,\sigma}(x)=\left(\frac{x\cdot \xi}{\mu R}\right)^{\sigma}=\exp\left[\sigma\log\left(\frac{x\cdot \xi}{\mu R}\right)\right] \label{solution}
\end{gather}
is a solution of \eqref{wave_eq} (written in the coordinate system presented in \eqref{dS}), where $\xi \in \sA$ is a~future directed null covector. $\sigma$ satisfies the equation{\samepage
\begin{gather}
\mu^{2}R^ {2} =-\sigma(n-1+\sigma) ,\label{sigma}
\end{gather}
where $n$ is the dimension of the de Sitter spacetime.}

We will search for real and complex solutions of \eqref{sigma}. The complex solution is
\begin{gather}
\operatorname{Re} \sigma = -\frac{n-1}{2} \qquad \textrm{and} \qquad \operatorname{Im} \sigma = \pm\sqrt{\mu^ {2}R^ {2}-(n-1)^{2}/4} \eqqcolon \pm\mu' .\label{mass_shell}
\end{gather}
In this case, the mass $\mu$ assumes a minimum value $\mu_{\min}=(n-1)/(2R)$ and $\pm\mu'$ can assume any real value. This solution corresponds to the so-called principal series of representations~\cite{Bargmann47} and describes a massive field on $\dS$.

The real solution of \eqref{sigma} is
\begin{gather*}
\sigma = -\frac{n-1}{2}+\mu'' \qquad \textrm{and} \qquad \mu'' = \pm\sqrt{(n-1)^ {2}/4-\mu^ {2}R^ {2}} .
\end{gather*}
Now, the mass is bounded from above and the corresponding Compton wavelength is of the order of the curvature radius. This corresponds to the so-called complementary series. These solutions present problems when one tries to interpret them as massive solutions of~\eqref{dalembert} \cite{AngFlaFronStern81,BarutBohm70,Garidi03}.

Henceforth we will concentrate on the principal series because, already on de Sitter spacetime, it provides a clearer interpretation of the mass. In this case, the plane waves indeed oscillate. However, \eqref{solution} is neither an eigenfunction of~${\bf m}^{2}$ nor of~${\bf j}^{2}$, although these operators commute. In addition, the determination of the mass $\mu$ does not impose any constraint on the covector~$\xi$,\footnote{Afterwards, when we contract the Lorentz group into the Poincar\'{e} group, the mass will actually constrain $\xi$.} but rather on the exponent~$\sigma$.

These solutions can be obtained by restricting the solutions of the massless wave equation on a subregion of $\M_{n+1}$ to $\dS$. This method leads directly to the Fourier transform on $\dS$ \cite{Strichartz73}. The procedure consists of some steps. Firstly a unitary operator between Hilbert spaces of even (or odd) functions on the light cone $\fC$ homogeneous of certain degree is constructed. Secondly this operator is extended to $\cL^2(\fC)$ (with volume measure) by using the Mellin transform. At this step one already obtains a Fourier transform between square integrable functions on the light cone. Thirdly one defines functions in the subregion of $\M_{n+1}$ described by \mbox{$\sqrt{\sum\limits_{j=1}^{n}(x_j)^2}>|x_0|$} using the Fourier expression of functions on the light cone, then proves that the original functions are boundary values of these newly defined ones and their restriction to~$\dS$ is surjective on~$\cL^2(\dS,\mrd\Sigma)$.

The region $\sqrt{\sum\limits_{j=1}^{n}(x_j)^2} \eqqcolon |{\bf x}|>|x_0|$ in $\M_{n+1}$ is covered by the hyperboloids
\begin{gather*}
\fH_R\colon \ \left\{p \in \M_{n+1} \,|\, -(x_0(p))^2+\sum_{k=1}^{n}(x_k(p))^2 = R^2 \right\} .
\end{gather*}
Let $\Box_R = \Box|_{\fH_R}$ and $f,g \in \cL^{2}(\fH_R,\mrd \Sigma)$ be such that $\Box_R f = \lambda^2 f$ and $\Box_R g = \lambda^2 g$. Then,
\begin{gather}
u(R,p) \coloneqq R^{-(n-1)/2 + \mri \rho}f(p) + R^{-(n-1)/2 - \mri \rho}g(p)\label{massless}
\end{gather}
is a solution of the massless wave equation on $\M_{n+1}$, $\Box u = 0$, where
\begin{gather*}
\rho = \sqrt{\lambda^2R^2 - (n-1)^2/4} .
\end{gather*}

Defining on $\fC$ the coordinates $s = |x_0|=|{\bf x}|$, ${\bf x'}=\frac{{\bf x}}{s}$, $x'_0=\frac{x_0}{s} = \pm 1$, if $v$ is a homogeneous function of degree $\sigma$ on $\fC$,
\begin{gather*}
v(s,x'_0,{\bf x'}) = s^\sigma\varpi(x'_0,{\bf x'}) ,
\end{gather*}
where $\varpi \in \cC^\infty\big({\pm}1 \times S^{n-1}\big)$. Completing in the topology given by the norm
\begin{gather*}
||f||_\sigma^2 \coloneqq \iint|\varpi(x'_0,{\bf x'})|^2\mrd x'_0\mrd^{n-1} x' ,
\end{gather*}
one obtains a Hilbert space $\cH_\sigma$. Let now $(\tau,{\bf \chi})$ be another point in $\fC$, $\zeta = |\tau|=|{\bf \chi}|$, ${\bf \chi'}=\frac{{\bf \chi}}{\zeta}$, $\tau'=\frac{\tau}{\zeta} = \pm 1$, then $(\tau,{\bf \chi})\cdot(t,{\bf x}) = (-\tau't'+{\bf \chi'}\cdot{\bf x'})s\zeta$. Let $a \coloneqq -\tau't'+{\bf \chi'}\cdot {\bf x'}$. The Fourier transform of a homogeneous function of degree $\sigma=-\frac{(n-1)}{2}+\mri\rho$ on $\fC$ is
\begin{gather}
Th(t,{\bf x}) = \int_0^\infty\!\iint h(\zeta,\tau',{\bf \chi'})\mre^{-\mri as\zeta}\zeta^{n-2}\mrd \zeta \mrd\tau'\mrd^{n-1}\chi' \nonumber \\
\hphantom{Th(t,{\bf x})}{}= \Upgamma\left(\frac{n-1}{2}+\mri\rho\right)s^{-\frac{(n-1)}{2} - \mri \rho}\mre^{-\frac{\mri\pi}{2}\left(\frac{n-1}{2}+\mri\rho\right)} \nonumber \\
\hphantom{Th(t,{\bf x})=}{}\times\iint|a|^{-\frac{n-1}{2}-\mri\rho}\left[\Theta(a)+\mre^{\mri\pi\left(\frac{n-1}{2}+\mri\rho\right)}\Theta(-a)\right] \varpi(\tau',{\bf \chi'})\mrd\tau'\mrd^{n-1}\chi' .\label{intertwiner}
\end{gather}
Since the Fourier transform preserves parity and the norm $||\cdot||_\sigma$ is invariant under the action of the group ${\rm SO}_0(1,n)$, the operator
\begin{gather*}
s^{-(n-1)/2 - \mri \rho}\iint|a|^{-\frac{n-1}{2}-\mri\rho}\left[\Theta(a)+\mre^{\mri\pi\left(\frac{n-1}{2}+\mri\rho\right)}\Theta(-a)\right]\times \cdot\mrd\tau'\mrd^{n-1}\chi'
\end{gather*}
is an intertwiner between equivalent representations of ${\rm SO}_0(1,n)$ on even (odd) functions in $\cH_\sigma$ and on even (odd) functions in $\cH_{-\sigma-n+1}$. Moreover, the regular representation of ${\rm SO}_0(1,n)$ acts irreducibly on the subspaces of even (odd) functions in $\cH_\sigma$. Hence, by Schur's lemma, the intertwiner is proportional to a unitary operator on each of the spaces $\cL^2_e\big({\pm}1 \times S^{n-1}\big)$ and $\cL^2_o\big({\pm} 1 \times S^{n-1}\big)$, respectively~$U_e$ and~$U_o$. But these spaces (as subspaces of $\cH_\sigma$) are orthogonal, therefore we obtain a unitary operator $\mathcal{U} \coloneqq U_e \oplus U_o$ on
\begin{gather*}\cL^2_e\big({\pm} 1 \times S^{n-1}\big) \oplus \cL^2_o\big({\pm}1 \times S^{n-1}\big) .\end{gather*}
The action of this operator can be explicitly determined by classical methods of Fourier analysis. We leave detailed calculations to Appendix~\ref{appendix}.

Denoting by $d(\rho)$ the constant of proportionality, the action of $\mathcal{U}$ on a function $\varpi \in \cC^\infty\big({\pm}1 \times S^{n-1}\big)$ is given by
\begin{gather}
(\mathcal{U}\varpi)(t',{\bf x'},\rho) = d(\rho)\!\iint|a|^{-\frac{n-1}{2}-\mri\rho}\left[\Theta(a)+\mre^{\mri\pi\left(\frac{n-1}{2}+\mri\rho\right)}\Theta(-a)\right]\!\times\! \varpi(\tau',{\bf \chi'})\mrd\tau'\mrd^{n-1}\chi' \nonumber \\
\hphantom{(\mathcal{U}\varpi)(t',{\bf x'},\rho)}{} \eqqcolon d(\rho)\psi(t',{\bf x'},\rho) .\label{intertw-unit}
\end{gather}
Therefore $\varpi(t',{\bf x'},\rho)=d(\rho)(\mathcal{U}^\ast\psi)(t',{\bf x'},\rho)$ is a smooth function in $\cC^\infty\big({\pm}1 \times S^{n-1}\big)$ depending on a real parameter $\rho$. Besides, if $h \in \cL^2(\fC)$, one can calculate the Mellin transform of this function:
\begin{gather*}
\varpi(t',{\bf x'},\rho) = \int_{0}^{\infty}h(t',{\bf x'},s)s^{\frac{n-1}{2}-\mri\rho}\frac{\mrd s}{s}, \\
h(t',{\bf x'},s) = \frac{1}{2\pi}\int_{-\infty}^{\infty}\varpi(t',{\bf x'},\rho)s^{-\frac{n-1}{2}+\mri\rho}\mrd \rho .
\end{gather*}
In the above equations, $s=|x_0|=|{\bf x}|$, but $\rho$ is a real parameter. Therefore $\varpi(t',{\bf x'},\rho)$ is a~smooth function in $\cC^\infty\big({\pm}1 \times S^{n-1}\big)$ depending on a real parameter $\rho$, as before. Therefore every function $h \in \cL^2(\fC)$ can be written in the form
\begin{gather}
 h(t,{\bf x}) = \frac{1}{2\pi}\int_{-\infty}^{\infty}\iint \psi(\tau',{\bf \chi'},\rho)|-t\tau'+{\bf x}\cdot{\bf \chi'}|^{-\frac{n-1}{2}+\mri\rho} \nonumber \\
\hphantom{h(t,{\bf x}) =}{} \times\left[\Theta(-t'\tau'+{\bf x'}\cdot{\bf \chi'})+\mre^{\mri\pi\left(-\frac{n-1}{2}+\mri\rho\right)}\Theta(t'\tau'-{\bf x'}\cdot{\bf \chi'})\right]|d(\rho)|^2\mrd\rho\mrd\tau'\mrd^{n-1}\chi'\label{F-inverse}
\end{gather}
and
\begin{gather}
 \psi(\tau',{\bf \chi'},\rho) = \int_{0}^{\infty}\iint h(t,{\bf x})|-t\tau'+{\bf x}\cdot{\bf \chi'}|^{-\frac{n-1}{2}-\mri\rho} \nonumber \\
\hphantom{\psi(\tau',{\bf \chi'},\rho) =}{} \times\left[\Theta(-t'\tau'+{\bf x'}\cdot{\bf \chi'})+\mre^{\mri\pi\left(\frac{n-1}{2}+\mri\rho\right)}\Theta(t'\tau'-{\bf x'}\cdot{\bf \chi'})\right]s^{n-2}\mrd s\mrd t'\mrd^{n-1}x' ,\label{F-transform}
\end{gather}
where
\begin{gather}
|d(\rho)| = (2\pi)^{-\frac{(n+1)}{2}}\frac{\left|\Upgamma\left(\frac{n-1}{2}+\mri\rho\right)\right|}{\left|\Upgamma(-\mri\rho)\right|} \nonumber \\
\hphantom{|d(\rho)| =}{} \times\begin{cases}
\pi\sqrt{2(1+\coth\pi\rho)} & \textrm{if $n$ is even}, \\
\pi\left(1+\tanh\dfrac{\pi\rho}{2}\right) & \textrm{if $n-1+2(j-k)$ is a multiple of 4}, \vspace{1mm}\\
\pi\left(1+\coth\dfrac{\pi\rho}{2}\right) & \textrm{if $n-1+2(j-k)$ is not a multiple of 4},
\end{cases} \label{d-parameter}
\end{gather}
where $j$ and $k$ are integer numbers such that, if $j+k$ is even, the function $\varpi$ is even, but if $j+k$ is odd, $\varpi$ is odd. The numbers $j$ and $k$ are defined in Appendix~\ref{appendix}. We remark that $\mrd\tau'$ is a discrete measure concentrated on $\{-1,1\}$. Thus $\int\mrd\tau' = \sum\limits_{\tau'=\pm 1}\tau'$.

This development corresponds to the first and second steps mentioned above. The third step starts from the observation made in~\cite{Strichartz73} that if one defines a function $u$ in the region $|{\bf x}| \geq |x_0|$ by the expression in~\eqref{F-inverse}, then $\Box u = 0$, $u|_{\fC} \in \cL^2(\fC)$ and if the integration in the real parame\-ter~$\rho$ is restricted either to $(0,\infty)$ or to $(-\infty,0)$, then $u|_{\fH_R} \in \cL^2(\fH_R)$. Furthermore, it is proved in~\cite{Strichartz73} that the map $u|_\fC \rightarrow (f,g) \in \cL^{2}(\fH_R)\times\cL^{2}(\fH_R)$ is onto, where~$u$, $f$ and $g$ are related as in equation~\eqref{massless}. Hence $f$ and $g$ are solutions of the massive wave equation on a hyperboloid such as~$\dS$.

Equation \eqref{F-transform} defines the Fourier transform of the function $h \in \cL^{2}(\dS,d\Sigma)$ and \eqref{F-inverse} is the inverse transform. Note that $\psi \in \cL^2(\fC)$. The Fourier transform gives a decomposition of $\cL^{2}(\dS,d\Sigma)$ into a direct integral of eigenspaces of $\Box_{\dS}$.

Returning to the previous notation, we note, as in \cite{BrosMoschella96}, that the integrations over $\pm 1 \times S^{n-1}$ may be substituted by integrations over any curve homotopic to those ones. Hence we identify $\tau' = \xi_0$, ${\bf \chi'} = {\bf \xi}$ and $\mrd\tau'\mrd^{n-1}\chi' \rightarrow \left[i_U\omega\right]$, where $\left[i_U\omega\right]$ is the contraction of the invariant volume form $\omega$ on $\fC$ with the vector $U$ tangent to a curve that intercepts each (or almost every) generator of $\fC$ exactly once.

In components,
\begin{gather*}
\omega=\frac{1}{2|\xi_0|}\mrd \xi_1\wedge \cdots \wedge \mrd \xi_n \qquad \textrm{and} \qquad U=(\xi_0, \ldots , \xi_n) .
\end{gather*}
Therefore,
\begin{gather}
\left[i_U\omega\right]=\frac{1}{2|\xi_0|}\sum_{j=1}^{n}(-1)^{j+1}\xi_j\mrd\xi_1 \wedge \cdots \wedge \widehat{\mrd\xi}_j \wedge \cdots \wedge \mrd\xi_n , \label{cone_measure}
\end{gather}
where $\widehat{\mrd\xi}_j$ means that the differential $\mrd\xi_j$ is not present in the product. Furthermore, we identify $t=x_0$ and write a plane wave as
\begin{gather}
\Psi_\mu(x,\xi)=\Theta(x\cdot\xi)\left|\frac{x\cdot\xi}{\mu R}\right|^{-(n-1)/2+\mri\mu'} \!+ \mre^{-\pi\left(\mri(n-1)/2+\mu'\right)}\Theta(-x\cdot\xi)\left|\frac{x\cdot\xi}{\mu R}\right|^{-(n-1)/2+\mri\mu'} . \!\!\!\label{psi_mu}
\end{gather}
The Heaviside functions (and the extra exponential factor) appear because the above expressions mix even and odd terms. Another interpretation of these terms is as reflecting a choice of branch cut in the definition of the logarithm, since $x\cdot\xi$ can be negative \cite{BrosEpsteinMoschella10,BrosMoschella96,BrosMoschella04}.

Alternatively, the plane waves can be written in the hyperbolic coordinate system defined in~\eqref{hyper_coord}. The d'Alembert operator becomes
\begin{gather*}
\Box_{\dS}=-\frac{\partial^2}{\partial \beta^2}-(n-1)\tanh\beta\frac{\partial}{\partial \beta}+\frac{\Delta}{\cosh^2\beta} ,
\end{gather*}
where $\Delta$ is the Laplace operator on $S^{n-1}$. The d'Alembert operator is thus separable. The Laplace operator on $S^{n-1}$ is expressed as
\begin{gather*}
\Delta = (\sin\varphi_1)^{-(n-2)}\frac{\partial}{\partial \varphi_1}\left((\sin\varphi_1)^{(n-2)}\frac{\partial}{\partial \varphi_1}\right) \nonumber \\
\hphantom{\Delta =}{} + (\sin\varphi_1)^{-2}(\sin\varphi_2)^{-(n-3)}\frac{\partial}{\partial \varphi_2}\!\left((\sin\varphi_2)^{(n-3)}\frac{\partial}{\partial \varphi_2}\right)\! + \cdots + (\sin\varphi_1 \cdots \sin\varphi_{n-2} )^{-2}\frac{\partial^2}{\partial\phi^2} .
\end{gather*}
The eigenfunctions of this differential operator can be written as products of associated Legendre functions:
\begin{gather*}
Y^{m}_{l_1,\ldots,l_{n-2}}(\varphi_1,\ldots,\varphi_{n-2},\phi) = C\!\left[\prod_{q=1}^{n-2}\left(\sin\varphi_{n-1-q}\right)^{-\left(\frac{q-1}{2}\right)}P_{l_q+\frac{q-1}{2}}^{l_{q-1}+\frac{q-1}{2}}(\cos\varphi_{n-1-q})\right]\!\mre^{\mri m\phi},
\end{gather*}
where $C$ is a normalization factor and $l_0 \coloneqq m$. Besides, $|m| \leq l_1 \leq \dots \leq l_{n-2}$ and all $l_q$ are positive integers. Moreover
\begin{gather*}
\Delta Y^{m}_{l_1,\ldots,l_{n-2}}(\varphi_1,\ldots,\varphi_{n-2},\phi) = -l_{n-2}(l_{n-2}+n-2)Y^{m}_{l_1,\ldots,l_{n-2}}(\varphi_1,\ldots,\varphi_{n-2},\phi) .
\end{gather*}

If a plane wave is written as
\begin{gather*}\psi(\beta,\varphi_1,\ldots,\varphi_{n-2},\phi)=V(\beta)Y^{m}_{l_1,\ldots,l_{n-2}}(\varphi_1,\ldots,\varphi_{n-2},\phi) ,
\end{gather*}
the function $V(\beta)$ satisfies the equation
\begin{gather*}
\left\{\frac{\partial^2}{\partial \beta^2}+(n-1)\tanh\beta\frac{\partial}{\partial \beta}+\rho^2 + \frac{(n-1)^2}{4}+\frac{l_1(l_1+n-2)}{\cosh^2\beta}\right\}V(\beta) = 0 ,
\end{gather*}
where $\rho$ is a nonnegatve real number. The solutions of this equation are given by gaussian hypergeometric functions. One thus obtains the two sets of plane waves:
\begin{gather}
 \tensor*[_1]{\Psi}{^{\rho,m}_{l_1,\ldots,l_{n-2}}}(\beta,\varphi_1,\ldots,\varphi_{n-2},\phi) = \frac{2}{\tensor[_1]{K}{^{1/2}}}Y^{m}_{l_1,\ldots,l_{n-2}}(\varphi_1,\ldots,\varphi_{n-2},\phi) \nonumber \\
 \times \, \tensor[_2]{F}{_1}\left(\frac{\mri\rho+l_{n-2}+\frac{(n+1)}{2}}{2},\frac{\mri\rho-l_{n-2}-\frac{(n-5)}{2}}{2};\frac{3}{2},\tanh^2\beta\right)\tanh\beta\left(\cosh\beta\right)^{-\frac{(n-1)}{2} + \mri\rho},\label{hyper-odd}
\\
 \tensor*[_2]{\Psi}{^{\rho,m}_{l_1,\ldots,l_{n-2}}}(\beta,\varphi_1,\ldots,\varphi_{n-2},\phi) = \frac{1}{\tensor[_2]{K}{^{1/2}}}Y^{m}_{l_1,\ldots,l_{n-2}}(\varphi_1,\ldots,\varphi_{n-2},\phi) \nonumber \\
 \times \, \tensor[_2]{F}{_1}\left(\frac{\mri\rho+l_{n-2}+\frac{(n-1)}{2}}{2},\frac{\mri\rho-l_{n-2}-\frac{(n-3)}{2}}{2};\frac{1}{2},\tanh^2\beta\right)\left(\cosh\beta\right)^{-\frac{(n-1)}{2} + \mri\rho} ,\label{hyper-even}
\end{gather}
where
\begin{gather*}
\tensor[_1]{K}{}=\frac{\pi\left[\cosh(\pi\rho)-(-1)^{l_{n-2}}\cos\left((n-1)\frac{\pi}{2}\right)\right]\left|\Upgamma\left[\frac{\mri\rho+l_{n-2}
+\frac{(n-1)}{2}}{2}\right]\right|^2}{\sinh(\pi\rho)\left|\Upgamma\left[\frac{\mri\rho+l_{n-2}+\frac{(n+1)}{2}}{2}\right]\right|^2}, \\
\tensor[_2]{K}{}=\frac{\pi\left[\cosh(\pi\rho)+(-1)^{l_{n-2}}\cos\left((n-1)\frac{\pi}{2}\right)\right]\left|\Upgamma\left[\frac{\mri\rho+l_{n-2} +\frac{(n+1)}{2}}{2}\right]\right|^2}{\sinh(\pi\rho)\left|\Upgamma\left[\frac{\mri\rho+l_{n-2}+\frac{(n-1)}{2}}{2}\right]\right|^2} .
\end{gather*}

The parity of the plane waves is determined by the factor $\alpha \in \{1,2\}$ and the highest angular momentum $l_{n-2}$: a plane wave is even (odd) if $\alpha + l_{n-2}$ is even (odd). These plane waves also form a dense subset of $\cL^2(\dS,\mrd\Sigma)$ and carry a unitary irreducible representation of the Lorentz group ${\rm SO}_0(1,n)$ \cite{LimNiedRacz66a,LimNiedRacz67,LimNiedRacz66b}. The Fourier transform of $f\in\cL^2(\dS,\mrd\Sigma)$ is given by
\begin{gather}
\tensor[_{1,2}]{\chi}{^{\rho,m}_{l_1,\ldots,l_{n-2}}}=\int_{\dS}\overline{\tensor*[_{1,2}]{\Psi}{^{\rho,m}_{l_1,\ldots,l_{n-2}}}} (\beta,\varphi_1,\ldots,\varphi_{n-2},\phi)f(\beta,\varphi_1,\ldots,\varphi_{n-2},\phi)\mrd\Sigma .\label{F-hyper}
\end{gather}
The terms $\tensor[_\alpha]{\chi}{^{\rho,m}_{l_1,\ldots,l_{n-2}}}$ form a square summable sequence, hence they span a Hilbert space. The inverse transform is given by
\begin{gather}
f(\beta,\varphi_1,\ldots,\varphi_{n-2},\phi) = \int_0^\infty \mrd\rho\sum_{\substack{\alpha,m,\\l_1,\ldots,l_{n-2}}}\tensor[_\alpha]{\chi}{^{\rho,m}_{l_1,\ldots,l_{n-2}}}\tensor*[_\alpha]{\Psi}{^{\rho,m}_{l_1,\ldots,l_{n-2}}}(\beta,\varphi_1,\ldots,\varphi_{n-2},\phi) .
\label{F-hyper-inverse}
\end{gather}

These two coordinate systems will be used in the analysis of wavepackets. The Fourier transform presented here is known in the literature as ``Fourier--Helgason transform''. It was originally formulated by Harish-Chandra (see \cite{Harish-Chandra_I-58,Harish-Chandra_II-58}) as a transformation on functions defined on a homogeneous space acted upon by a semi-simple Lie group (as the Lorentz group). Later it received a more geometric interpretation by Helgason (see~\cite{Helgason68,Warner-II}). We will call it here simply ``Fourier transform''.

\section{Contraction of the Lorentz algebra}\label{sec_contract}

After we construct the wavepackets on $\dS$ we will compare them with the wavepackets constructed on~$\M_n$~\cite{Hepp65} (note that $\M_n$ has the same number of dimensions as $\dS$, differently from~$\M_{n+1}$, in which $\dS$ is embedded). Although the spacetimes are different, at any point of a curved spacetime (of dimension $n$) the components of the metric tensor can be made equal to the components of the metric tensor of $\M_n$ \cite{ONeill83}, hence a (possibly infinitesimal) neighbourhood of the curved spacetime resembles (i.e., the effects of the curvature can be treated as perturbations) the Minkowski spacetime. Besides, on the level of Lie algebras, the contraction of algebras~\cite{InonuWigner53} allows us to transform the Lorentz algebra $\mathfrak{so}(1,n)$ into the Poincar\'{e} algebra $\fp_n \coloneqq\mathfrak{so}(1,n-1)\subsetplus\mathbb{R}^{n}$.

As observed in \cite{Hannabuss71}, almost every element $g\in L$ can be written as the product $g=\text{ban}$, with $ \text{b}\in B$, where $B$ is the normalizer of a subgroup of $L$ isomorphic to $\text{SO}_0(1,n-1)$. $B$ is composed of elements of the form
\begin{gather*}
B \coloneqq \begin{pmatrix}
\text{SO}_0(1,n-1) & 0 \\
0 & \pm 1
\end{pmatrix} .
\end{gather*}

Let now $b \in \fb$ (the Lie algebra associated to $B$), $a \in \fa$ and $n_i \in \fn$ and define
\begin{gather}
b' \coloneqq b , \qquad a' \coloneqq \frac{1}{R}a \qquad \text{and} \qquad n'_i \coloneqq \frac{1}{R}n_i .\label{alg_contract}
\end{gather}
In the limit $R \rightarrow \infty$, one verifies that the commutation relations satisfied by $b'$, $a'$ and $n'$ are those satisfied by the generators of the Poincar\'{e} algebra $\fp_n$, where the elements $b'$ generate the subalgebra $\mathfrak{so}(1,n-1)$ (as $b$ did), $a'$ is the generator of time translations and $n'$ are the generators of spatial translations. $a'$ and $n'$ comprise the abelian subalgebra which is normalized by $\mathfrak{so}(1,n-1)$.

The two pictures above are comparable. On the de Sitter spacetime, one may consider a~neighbourhood $\mathcal{O}_p$ of a point $p\in\dS$ such that $\forall\, x\in\mathcal{O}_p$, the components of the metric tensor in the region $\mathcal{O}_p$ are equal to the components of the metric tensor of the Minkowski spacetime~$\M_n$. Hence, after the limit $R \rightarrow \infty$ is taken, the spacetime becomes the Minkowski spacetime~$\M_n$ and its group of isometries is the Poincar\'{e} group~$P_n$.

\subsection{Representation of the Poincar\'{e} algebra}\label{subsec_poinc}

The representation of the Lorentz group defined in~\eqref{group_repr} can be induced from a representation of the closed subgroup
\begin{gather*}
Q=MAN \subset L .
\end{gather*}
Moreover, the induction procedure shows that the functions $f\in\cL^2(L,\mrd L)$ are completely determined if their values on the quotient space $BAN/MAN=B/M$ are known (see the construction of the induced representation in \cite[Section 6.1]{Folland15}), where $\mrd L$ is the unique left invariant Haar measure on $L$. However, $B/M$ is diffeomorphic to the hyperboloids \cite{Hannabuss71,MickNied72}
\begin{gather*}
-x_{0}^{2}+x_{1}^{2}+\dots +x_{n-1}^{2}=-R^{2} ,
\end{gather*}
which are the intersections between the hyperplane $x_n^2=2R^2$ and the de Sitter spacetime:
\begin{gather*}
\dS_b^\pm \coloneqq \big\{p\in\dS \,|\, {-}x_{0}(p)^{2}+x_{1}(p)^{2}+\dots +x_{n-1}(p)^{2}=-R^{2} ,\, x_0(p) \gtrless 0\big\} .
\end{gather*}

The quotient space $B/M$ remains unaltered by the contraction procedure. Therefore, after the contraction, the representation of the Poincar\'{e} group $P_n$ is determined by its representations on the two disjoint hyperboloids $\dS_b^+$ and $\dS_b^-$, i.e., it is the direct sum of these subrepresentations. The Hilbert spaces on which these subrepresentations act are $\cL^2(\dS_b^+,\mrd\Sigma)$ and $\cL^2(\dS_b^-,\mrd\Sigma)$, respectively, which are subspaces of $\cL^2(\dS,\mrd\Sigma)$. Therefore the identity operator $\mathbbm{1}$ on $\cL^2(\dS,\mrd\Sigma)$ is also an identity operator on both $\cL^2(\dS_b^+,\mrd\Sigma)$ and $\cL^2(\dS_b^-,\mrd\Sigma)$.

On the other hand, the irreducible representation on $\cL^2(\dS_b^+,\mrd\Sigma)$ differs from the irreducible representation on $\cL^2(\dS_b^-,\mrd\Sigma)$ only by the change $\mu' \rightarrow -\mu'$ \cite{MickNied72}, where $\mu'$ is related to the mass of the plane wave of~$\dS$ by~\eqref{mass_shell}. Furthermore, the mass $\mu$ is related to the eigenvalue of the Casimir operator ${\bf C}^2$ of the Lorentz algebra $\frl$. As we will see now, the eigenvalues of this operator can be easily related to the eigenvalues of a Casimir operator ${\bf \cP}^2$ of the Poincar\'{e} algebra $\fp_{n}$.

Let us focus on the irreducible representation of the Lorentz algebra presented in~\eqref{m_i0-ij}, restricted to, say, $\cL^2(\dS_b^+,\mrd\Sigma)$. Both this representation and the one arising after contraction are irreducible. The Casimir operator is ($1\leq i,j \leq n-1$)
\begin{gather*}
\sum_{i<j}{\bf m}_{ij}{\bf m}_{ij}+\sum_{i}{\bf m}_{in}{\bf m}_{in}-\sum_{i}{\bf m}_{i0}{\bf m}_{i0}-{\bf a}^{2}=-\mu^ {2}R^{2}\mathbbm{1} ,
\end{gather*}
where ${\bf a}$ is the representation of $a=m_{n0}$. Denoting the representations of $n_i$ by ${\bf n}_i$,
\begin{gather}
\sum_{i<j}{\bf m}_{ij}{\bf m}_{ij}+\sum_{i}({\bf n}_{i}-{\bf m}_{i0})({\bf n}_{i}-{\bf m}_{i0})-\sum_{i}{\bf m}_{i0}{\bf m}_{i0}-{\bf a}^{2}= -\mu^ {2}R^{2}\mathbbm{1} \therefore , \nonumber \\
\frac{1}{R^2}\sum_{i<j}{\bf m}_{ij}{\bf m}_{ij}+\sum_{i}\left({\bf n'}_{i}{\bf n'}_{i}-{\bf n'}_{i}\frac{1}{R}{\bf m}_{i0}-\frac{1}{R}{\bf m}_{i0}{\bf n'}_{i}\right)-({\bf a'})^{2}= -\mu^ {2}\mathbbm{1} .\label{dalembertian-bef_limit}
\end{gather}
Hence, after we take the limit $R\rightarrow\infty$, we find the Casimir operator ${\bf \cP}^2$ of the Poincar\'{e} algebra~$\fp_{n}$, together with its eigenvalue:
\begin{gather}
\sum_{i}({\bf n'}_i)^{2}-({\bf a'})^{2}=-\mu^ {2}\mathbbm{1} \eqqcolon {\bf \cP}^2 .\label{dalembertian-limit}
\end{gather}
If we had restricted the irreducible representation of the Lorentz group to the Hilbert space $\cL^2(\dS_b^-,\mrd\Sigma)$, we would only need to make the modification $\mu' \mapsto -\mu'$, and the above reasoning would apply without further changes. But this modification does not change the value of~$\mu^ {2}$. Therefore the eigenvalue of the Casimir operator ${\bf \cP}^2$ of the Poincar\'{e} algebra $\fp_n$ is uniquely determined by the eigenvalue of the Casimir operator ${\bf C}^2$ of the Lorentz algebra $\frl$. This result is the first step in the comparison of the wavepackets on $\dS$, to be constructed in the following section, and the usual wavepackets on~$\M_n$. We record this result in the following

\begin{Lemma}\label{lemma_contract}The irreducible representation of the Casimir operator ${\bf C}^2$ of the Lorentz algebra~$\frl$, with eigenvalue $-\mu^2R^2$, denoted by $_{\mu}{\bf C}^2$, under the contraction of the Lorentz algebra $\frl$ into the Poincar\'{e} algebra $\fp_n$ presented in the former section, is contracted towards the direct sum of two irreducible representations of the Casimir operator ${\bf \cP}^2$ of the Poincar\'{e} algebra $\fp_{n}$, both of them with eigenvalue $-\mu^2$, denoted by $_{\pm \mu}{\bf \cP}^2$,
\begin{gather*}
_{\mu}{\bf C}^2 \underset{R \rightarrow \infty}{\longrightarrow} {_{+\mu}{\bf \cP}^2} \oplus {_{-\mu}{\bf \cP}^2} .
\end{gather*}
\end{Lemma}

\section{Wavepackets}\label{sec_wpacket}

Recalling the Fourier transform presented in Section~\ref{sec_sols}, we are able to define

\begin{Definition}\label{def_wpack}
A wavepacket $f$ is the inverse Fourier transform of $\hat{f}\delta_{\mu'}$, where $\hat{f}$ is a compactly supported smooth function on the absolute $\sA$ (defined in Section~\ref{sec_def}) and $\delta_{\mu'}$ is the compactly supported Dirac distribution acting on functions on $\mathbb{R}$:
\begin{gather}
f(x)=|d(\mu')|^2\int_{\sA} \hat{f}(\xi)\Psi_{\mu}(x,\xi)\mrd\sA .\label{wpack}
\end{gather}
$\mrd\sA$ is the measure given in equation~\eqref{cone_measure}. In the hyperbolic coordinate system, a wavepacket is written as (see~\eqref{F-hyper-inverse})
\begin{gather}
f(\beta,\varphi_1,\ldots,\varphi_{n-2},\phi) = \sum_{\substack{\alpha,m,\\l_1,\ldots,l_{n-2}}} \tensor[_\alpha]{\chi}{^{\mu',m}_{l_1,\ldots,l_{n-2}}}\tensor*[_\alpha]{\Psi}{^{\mu',m}_{l_1,\ldots,l_{n-2}}}(\beta,\varphi_1,\ldots,\varphi_{n-2},\phi) .\label{wpack-hyper}
\end{gather}
\end{Definition}

We note the following important property of a wavepacket on $\dS$: the restriction of a~wave\-packet to a Cauchy hypersurface of $\dS$ is a smooth function. This is immediate from expression~\eqref{wpack-hyper} and is also a property of the wavepackets usually defined on~$\M_{n+1}$.

The restriction of the measure on the cone to the measure on the absolute (which is actually a~restriction from~$\fC$ to~$\fC^+$) will be justified when we calculate the behaviour of a wavepacket in the limit $R \rightarrow \infty$. Moreover, $\hat{f}\Psi_\mu$ is a bounded and infinitely differentiable function (for the expression in the ambient coordinate system, this is true in the regions where the sign of~$x\cdot\xi$ does not change). Then, from the dominated convergence theorem, the limit and integration operations commute, hence the derivatives of the wavepacket are the smearing of~$\hat{f}$ with the derivatives of the plane wave. Therefore the wavepacket is also a solution of the wave equation~\eqref{wave_eq}.

\subsection{Asymptotic behaviour}\label{subsec_asymptotic}

The estimate of the asymptotic behaviour of the wavepacket is crucial for the construction of a state which, asymptotically, can be interpreted as a free particle. The wavepacket on~$\dS$ is defined in~\eqref{wpack} above. Since $\xi\in \sA$, $\xi\cdot\xi=0$ and $\xi_0=\sqrt{\sum\limits_{i=1}^{n}(\xi_i)^{2}}$, the wavepacket is a function of fast decrease, in the sense that, if $s\coloneqq |x_0|+|{\bf x}|\rightarrow \infty$, where $|{\bf x}| \coloneqq \sqrt{\sum\limits_{i=1}^{n}(x_i)^2}$, the wavepacket decays faster than any inverse power of~$s$. This is stated and proved in the following

\begin{Theorem}\label{wpack-fast}A wavepacket on de Sitter spacetime is a function of fast decrease.
\end{Theorem}
\begin{proof}First, we prove the theorem using the ambient coordinate system.

Since the derivatives of the wavepacket are the smearing of $\hat{f}$ with the derivatives of the plane wave, we will focus only on $\Psi_\mu$ (see equation \eqref{psi_mu}). We will apply the stationary phase method to estimate the asymptotic behaviour of~$f$~\cite{SteinShakarchi-IV}. The oscillatory terms are
\begin{align*}
\left|\frac{x\cdot\xi}{\mu R}\right|^{-(n-1)/2+\mri\mu'} &=\exp\left\{\left(-\frac{n-1}{2}+\mri\mu'\right)\log\left|\frac{-x_0\sqrt{\sum\limits_{i=1}^{n}(\xi_i)^{2}}+\sum\limits_{i=1}^{n}x_i\xi_i}{\mu R}\right|\right\} \\
&=\exp\left\{(|x_0|+|{\bf x}|)\left(-\frac{n-1}{2}+\mri\mu'\right)\Phi_x(\xi)\right\} ,
\end{align*}
where
\begin{gather}
\Phi_x(\xi) \coloneqq \frac{1}{|x_0|+|{\bf x}|}\log\left|\frac{-x_0\sqrt{\sum\limits_{i=1}^{n}(\xi_i)^{2}}+\sum\limits_{i=1}^{n}x_i\xi_i}{\mu R}\right| .
\label{phase}
\end{gather}
 It must first be checked whether $\Phi_{x}(\xi)$ has any fixed points.
\begin{gather*}
\grad \Phi_x(\xi)=\frac{1}{(|x_0|+|{\bf x}|)(x\cdot\xi)}\left(x_1-x_0\frac{\xi_1}{\xi_0}, \ldots , x_n-x_0\frac{\xi_n}{\xi_0}\right)
\end{gather*}
(this expression is valid for the two possible signs of $x\cdot\xi$). However, $\forall\, i\in \{1,\ldots, n \}$, $-1\leq \frac{\xi_i}{\xi_0}\leq 1$. Besides, $(x_0)^{2}+R^{2}=\sum\limits_{i=1}^{n}(x_i)^{2}$. Therefore, if $\forall\, i\in \{1,\ldots, n \}$, $x_0\frac{\xi_i}{\xi_0}=x_i$,
\begin{gather*}
(x_0)^{2}+R^{2}=\left(\frac{x_0}{\xi_0}\right)^{2}\sum_{i=1}^{n}(\xi_i)^{2}=(x_0)^{2} \therefore R=0 ,
\end{gather*}
which is an absurd. Therefore $\grad \Phi_x(\xi) \neq 0$, hence the stationary phase method tells that, for~$s$ sufficiently large and $\forall\, m>0$, $\exists\, c>0$ such that
\begin{gather*}
f(x)\leqslant c (1+s )^{-m} .
\end{gather*}

The proof in the hyperbolic coordinate system is also instructive. In this case, $s=\mre^{|\beta|}$, hence we will calculate the limit of a wavepacket as $\beta \rightarrow \infty$ (for $\beta \rightarrow -\infty$ the result is the same).

The odd and even plane waves \eqref{hyper-odd}, \eqref{hyper-even} have a similar asymptotic behaviour. First we note that as $\beta \rightarrow \infty$ the argument of the hypergeometric functions converges to~1. We thus employ the following identity for hypergeometric functions~\cite{AndrewsAskeyRoy00}
\begin{gather*}
\tensor[_2]{F}{_1}(a,b;c,v) = \frac{\Upgamma(c)\Upgamma(c-a-b)}{\Upgamma(c-a)\Upgamma(c-b)}\tensor[_2]{F}{_1}(a,b;a+b+1-c,1-v) \\
\hphantom{\tensor[_2]{F}{_1}(a,b;c,v) =}{} +\frac{\Upgamma(c)\Upgamma(a+b-c)}{\Upgamma(a)\Upgamma(b)}(1-v)^{c-a-b}\tensor[_2]{F}{_1}(c-a,c-b;1+c-a-b,1-v) ,
\end{gather*}
which is valid since none of the entries of the Gamma and hypergeometric functions are negative integers (neither zero). But since $v=\tanh^2\beta$, $1-v=\sech^2\beta$ and $\lim\limits_{\beta\rightarrow\infty}\sech^2\beta = 0$. Besides, in both cases, $c-a-b=-\mri\rho$. Therefore, $\lim\limits_{\beta\rightarrow\infty}\tensor[_2]{F}{_1}\big(a,b;c,\tanh^2\beta\big)=D\cosh^{2\mri\rho}\beta$, where $D$ is a~constant.

Moreover, $\lim\limits_{\beta\rightarrow\pm\infty}\tanh\beta = \pm 1$ hence the leading term of both even and odd plane waves is
\begin{gather*}D' (\cosh\beta )^{-\frac{(n-1)}{2} + 3\mri\rho}Y^{m}_{l_1,\ldots,l_{n-2}}(\varphi_1,\ldots,\varphi_{n-2},\phi) .\end{gather*}
We may consider, without loss of generality, $\varphi_1 = \dots = \varphi_{n-2} = 0$ and approximate $\cosh\beta \approx \mre^\beta$, hence the leading term becomes
\begin{gather*}D'\mre^{-\frac{(n-1)}{2}\beta + \mri[3\rho\beta + m\phi]} .\end{gather*}
Since $\rho$ and $m$ are independent of each other, the oscillatory phase method gives the same result as before. Moreover, the leading term has an exponentially decreasing factor. Therefore, the conclusion is maintained.
\end{proof}

The asymptotic behaviour of the wavepacket on de Sitter spacetime is different from the analogue case on Minkowski spacetime \cite{Hepp65,Ruelle62}. There the phase of the wavepacket had critical points located along a trajectory in spacetime whose tangent vector was the phase velocity of the packet. Inside a neighbourhood of that trajectory the amplitude of the wavepacket decayed at a certain rate. Outside of that neighbourhood, the amplitude decreased fast. In the present case there is no stationary point. This is characteristic of harmonic analysis on semi-simple Lie groups \cite{Gangolli71,Harish-Chandra_II-58,Helgason66,Warner-II}.

\subsection{Flat limit}\label{subsec_flat}

At last we want to compare the behaviour of the wavepacket \eqref{wpack} in the limit $R \rightarrow \infty$ with the usual construction performed on the Minkowski spacetime $\M_n$. Besides showing consistency of our construction, this comparison will allow us to find an interpretation for $\xi$.

The plane wave \eqref{psi_mu} exhibits two terms, one that is different from zero only at the points satisfying $x\cdot\xi >0$, and another term, which is nonzero only for $x\cdot\xi < 0$. This second term contains a multiplicative factor that converges to zero exponentially fast in the limit $R\rightarrow\infty$. Therefore in the following we will analyse the behaviour of the wavepacket in the region where $x\cdot\xi >0$.

The function $\hat{f}$ is compactly supported on $\sA$. We consider its support to be contained in a~small neighbourhood of a covector~$\xi'\in\sA$. If we choose a coordinate system in which $\xi'$ has components $ (1,0,\ldots,0,1 )$, then
\begin{gather*}
x\cdot\xi'>0 \ \Rightarrow \ x_n>x_0 ,
\end{gather*}
which corresponds to a half of $\dS$ which will be designated as $\dS_n^+$. If, however, we pick a~different element $\xi''$ in the support of $\hat{f}$ and choose a new coordinate system such that, now, $\xi''$ has those same components, then
\begin{gather*}
x\cdot\xi''>0 \ \Rightarrow \ x'_n>x_0 .
\end{gather*}
Since the elements of $\sA$ can be obtained one from another by a rotation, the half of $\dS$ characterized by $x'_n>x_0$ is just a rotation of $\dS_n^+$. Therefore the region where $x\cdot\xi>0$ for every $\xi \in \mathrm{supp} \hat{f}$, the intersection of all the regions just described, is contained in $\dS_n^+$. On the other hand, the region where $x\cdot\xi>0$ for some $\xi \in \mathrm{supp}\hat{f}$ comprises a neighbourhood containing $\dS_n^+$. But since $\mathrm{supp}\hat{f}$ is contained in a small neighbourhood of $\xi'$, the regions where $x\cdot\xi>0$ for some $\xi \in \mathrm{supp} \hat{f}$ and $x_n<x_0$ are contained in small neighbourhoods of $x_n=x_0$.

We are first going to analyse the asymptotic behaviour of the plane wave in $\dS_n^+$ using the chart described in~\eqref{horo_coord}, with $\varepsilon=+1$.

The contraction of the Lorentz algebra presented in Section~\ref{sec_contract} showed that the generators of the horospheric translations contract into the generators of spatial translations and $a$ contracts into the generator of time translations. Moreover, \eqref{horo_coord} gives the change of coordinates from $(x_0, \ldots, x_n)$ to $(\tau, {\bf y})$. Actually, this is a restriction to a submanifold, from $\M_{n+1}$ to $\dS_n^+$. This fact will play an important role in the following. Analysing~\eqref{horo_coord} we find
\begin{gather*}
\frac{x_n-x_0}{R} = \mre^ {-\tau/R} , \qquad \mathrm{therefore} \qquad \begin{cases}
\tau = -R\log\left(\dfrac{x_n-x_0}{R}\right) \quad \textrm{and} \vspace{1mm}\\
y_i = -\dfrac{Rx_i}{x_n-x_0}.
\end{cases}
\end{gather*}
These equations are necessary in order to write the partial derivatives in the new coordinate system. The generators $a$ and $n_i$ are represented as differential operators as (see~\eqref{m_i0-ij} and~\eqref{m-n})
\begin{gather*}
{\bf a}= \mri R\left(\frac{\partial}{\partial\tau}+\sum_{i}\frac{y_i}{R}\frac{\partial}{\partial y_i}\right) \qquad \textrm{and} \qquad {\bf n}_i= \mri R\frac{\partial}{\partial y_i} .
\end{gather*}
We remark here that the horospheric translation is represented simply as a partial differentiation. Moreover, all the other terms in the differential operator \eqref{dalembertian-bef_limit} have higher powers of $1/R$.

Now, we are going to act with the operators ${\bf n'}_i{\bf n'}_i$ ($1 \leq i \leq n-1$) and ${\bf a'}^2$ (defined in~\eqref{alg_contract}) on the plane wave $\psi_\mu$ and analyse the behaviour of the result in the limit $R \rightarrow \infty$.
\begin{gather*}
{\bf n'}_i{\bf n'}_i \left(\frac{x\cdot \xi}{\mu R}\right)^{\sigma} = -\frac{\sigma}{\mu R} \left(\frac{x\cdot \xi}{\mu R}\right)^{\sigma-2} \nonumber \\
\hphantom{{\bf n'}_i{\bf n'}_i \left(\frac{x\cdot \xi}{\mu R}\right)^{\sigma} =}{}
\times \left\{\frac{\sigma-1}{\mu R}\left[\frac{y_i}{R}(\xi_0-\xi_n)-\xi_i\right]^{2}\mre^{-2\tau/R}+\left(\frac{x\cdot \xi}{\mu R}\right)\frac{\xi_0-\xi_n}{R}\mre^{-\tau/R}\right\} ,
\end{gather*}
and
\begin{gather*}
{\bf a'}^2\left(\frac{x\cdot \xi}{\mu R}\right)^{\sigma} = -\frac{\sigma}{\mu R}\left(\frac{x\cdot \xi}{\mu R}\right)^{\sigma-2} \nonumber \\
\hphantom{{\bf a'}^2\left(\frac{x\cdot \xi}{\mu R}\right)^{\sigma} =}{} \times \left\{\frac{\sigma-1}{\mu R}\left[-\cosh\left(\frac{\tau}{R}\right)\xi_0+\sinh\left(\frac{\tau}{R}\right)\xi_n+\frac{1}{2}\frac{|y|^{2}}{R^{2}}\mre^{-\tau/R}(\xi_0-\xi_n)\right]^{2} \right. \\
\left.\hphantom{{\bf a'}^2\left(\frac{x\cdot \xi}{\mu R}\right)^{\sigma} =}{}
+ \left(\frac{x\cdot \xi}{\mu R}\right)\!\left[-\sinh\left(\frac{\tau}{R}\right)\frac{\xi_0}{R}+\cosh\left(\frac{\tau}{R}\right)
\frac{\xi_n}{R}+\frac{1}{2}\frac{|y|^{2}}{R^{2}}\mre^{-\tau/R}\frac{(\xi_0-\xi_n)}{R}\right]\!\right\}.
\end{gather*}

Yet,
\begin{gather*}
\frac{\sigma}{\mu R}=-\frac{n-1}{2\mu R}+\mri\sqrt{1-\left(\frac{n-1}{2\mu R}\right)^2}=\mri+\mO(1/R)
\end{gather*}
and the same is valid for $\frac{\sigma-1}{\mu R}$. Hence
\begin{gather}
{\bf n'}_i{\bf n'}_i \left(\frac{x\cdot \xi}{\mu R}\right)^{\sigma} = \left(\frac{x\cdot \xi}{\mu R}\right)^{\sigma-2}\left[(\xi_i)^2 +\mO(1/R)\right] \text{and} \label{nn-limit} \\
{\bf a'}^2 \left(\frac{x\cdot \xi}{\mu R}\right)^{\sigma} = \left(\frac{x\cdot \xi}{\mu R}\right)^{\sigma-2}\left[(\xi_0)^2 +\mO(1/R)\right] . \label{aa-limit}
\end{gather}
Besides, $x\cdot\xi=-\tau\xi_0+\sum_i y_i(-\xi_i)+R\xi_n+\mO(1/R)$, hence
\begin{gather*}
\frac{x\cdot \xi}{\mu R}=\frac{-\tau\xi_0+\sum_i y_i(-\xi_i)}{\mu R}+\frac{\xi_n}{\mu}+\mO\big(1/R^2\big) .
\end{gather*}
Since the coordinates of a point of $\dS_n^+$ are parametrized by the pair $(\tau,{\bf y})$, we define $y=(\tau,{\bf y})$, $\overline{\xi}=(\xi_0,-\xi_i)$ and $y\cdot\overline{\xi} \coloneqq -\tau\xi_0+\sum_i y_i(-\xi_i)$. Therefore
\begin{gather*}
\left(\frac{x\cdot \xi}{\mu R}\right)^{\sigma-2} =\left(\frac{y\cdot\overline{\xi}}{\mu R}+\frac{\xi_n}{\mu}\right)^{\mu R(\mri+\mO(1/R))} =\left(\frac{y\cdot\overline{\xi}}{\xi_n R}+1\right)^{\xi_n R\left(\mri\frac{\mu}{\xi_n}+\mO(1/R)\right)}\left(\frac{\xi_n}{\mu}\right)^{\mu R(\mri+\mO(1/R))}.
\end{gather*}
If $\xi_n\neq\mu$, by the principle of stationary phase, for large values of $R$, the term
\begin{gather*}
\left(\frac{\xi_n}{\mu}\right)^{\mu R(\mri+\mO(1/R))}
\end{gather*}
is a fast decreasing function of $R$. Therefore
\begin{gather*}
\lim_{R\rightarrow\infty}\left(\frac{x\cdot \xi}{\mu R}\right)^{\sigma-2}=0 .
\end{gather*}
However, if $\xi_n=\mu$,
\begin{gather}
\lim_{R\rightarrow\infty}\left(\frac{x\cdot \xi}{\mu R}\right)^{\sigma-2}=\lim_{R\rightarrow\infty}\left(\frac{y\cdot\overline{\xi}}{\mu R}+1\right)^{\mu R(\mri+\mO(1/R))}=\mre^{\mri y\cdot\overline{\xi}}\label{wave-limit}
\end{gather}
and
\begin{gather}
\overline{\xi}\cdot\overline{\xi}=-\mu^ 2 .
\label{mass_shell-Mink}
\end{gather}
Note that, if the exponent was modified from $\sigma -2$ to $\sigma$, the above results would remain unchanged.

Therefore, collecting the results \eqref{wave-limit}, \eqref{aa-limit} and \eqref{nn-limit} and considering the action of the operator \eqref{dalembertian-limit} on the plane wave, one finds
\begin{gather*}
\lim_{R\rightarrow\infty}\left[\sum_{i}({\bf n'}_i)^{2}-({\bf a'})^{2}\right]\left(\frac{x\cdot \xi}{\mu R}\right)^{\sigma}=\big[(\xi_i)^2 - (\xi_0)^2\big]\mre^{\mri y\cdot\overline{\xi}}=-\mu^2\mre^{\mri y\cdot\overline{\xi}} .
\end{gather*}
$y$ represents the coordinates of a point of the resulting Minkowski spacetime $\M_n$ and $\overline{\xi}$ is a~timelike vector on the mass shell of~$\M_n$~\eqref{mass_shell-Mink}. Their origin, however, are restrictions of the coordinates~$x$ of a point on~$\dS$ and the null covector $\xi$ on the absolute, respectively. All other terms of $\Box_{dS}$ have higher powers of~$1/R$, and therefore their contribution would converge to zero.

We still have to analyse the asymptotic behaviour of $\psi_\mu$ in the points where $x\cdot\xi>0$ for some $\xi \in \mathrm{supp}\hat{f}$, but $x_n<x_0$. This region can be covered by the chart~\eqref{horo_coord}, with $\varepsilon=-1$. Since those points are contained in a small neighbourhood of $x_n=x_0$, the previous analysis shows that, after the limit $R\rightarrow\infty$, these points will be at infinity. Thus they are irrelevant, because of the decay of the wavepacket. Therefore we have proved that the plane wave on the whole~$\M_n$ is the limit $R\rightarrow\infty$ of the plane wave on half of $\dS$. Collecting these results in a~sentence, on~$\dS_n^+$,
\begin{gather*}
\lim_{R\rightarrow\infty}\psi_\mu(x,\xi)=\begin{cases}
\mre^{\mri y\cdot\overline{\xi}}, & \textrm{if $\xi_n=\mu$}, \\
0, &\textrm{if $\xi_n\neq\mu$}.
\end{cases}
\end{gather*}
This function is supported on a set of measure zero in the variable $\xi_n$. Hence when we multiply this limit with $\hat{f}$ and integrate on the absolute with the measure \eqref{cone_measure}, the integration over
\begin{gather*}
\frac{1}{2|\overline{\xi}_0|}\mrd\overline{\xi}_1\wedge \cdots \wedge \mrd\overline{\xi}_{n-1}
\end{gather*}
is the only one that gives a nonzero result.

Before collecting the results of this section, let us note that the plane wave and wavepacket on~$\dS$ are of fast decay, but this is not the behaviour of a wavepacket on Minkowski space\-time~\mbox{\cite{Hepp65,Hepp_Brandeis65}}. We can see the change in the asymptotic behaviour as~$R$ becomes larger. In~\eqref{wave-limit} one notes that the leading term in the exponent of the plane wave (above~\eqref{phase}), for large $R$, is
\begin{gather*}
\frac{1}{s}\mu R\log\left[1+\frac{y\cdot\overline{\xi}}{\mu R}\right] \eqqcolon \Upgamma_y(\overline{\xi})
\end{gather*}
(the term $i$ is not important now). Hence if we subtract and add this term to the phase,
\begin{gather*}
\left[\mu R\Phi_x(\xi)-\Upgamma_y(\overline{\xi})\right]+\Upgamma_y(\overline{\xi}) ,
\end{gather*}
the term between brackets goes to zero in the limit $R\rightarrow\infty$, but $\grad \Upgamma_y(\overline{\xi})$ has fixed points. Hence as the de Sitter radius~$R$ gets larger, the wavepacket decreases more slowly, until it reaches the rate~$s^{-3/2}$ given by the stationary phase principle~\cite{SteinShakarchi-IV} and calculated in~\cite{Hepp65, Hepp_Brandeis65}.

All the results of this section are collected in the following

\begin{Theorem}\label{wpack-flat}The limit, as $R\rightarrow\infty$, of the wavepacket on the $n$-dimensional de Sitter spacetime is a wavepacket on Minkowski spacetime, analytic in the whole $\M_n$, with mass sharply constrained to the mass shell $($now in the sense of~\eqref{mass_shell-Mink}$)$ and determined by the mass of its precedent wavepacket on $\dS$.
\end{Theorem}

\section{Conclusions}\label{sec_concl}

We have proved that one can consistently construct wavepackets on the de Sitter spacetime whose mass is defined from one of the Casimir elements in the universal enveloping algebra of the Lorentz algebra. The wavepacket is a function of fast decrease, differently from the wavepacket defined on Minkowski spacetime in \cite{Hepp65,Hepp_Brandeis65}. As we emphasized before, this is a~general feature of harmonic analysis on semi-simple Lie groups. The physical interpretation of the wavepacket became clearer after the evaluation of its flat limit, with the wavepackets converging to the usual one defined on Minkowski spacetime with support sharply constrained on the mass shell.

Besides being the first time that wavepackets are constructed on de Sitter spacetime, this is the first time that such a construction is made on any curved spacetime. The extension of this result to other curved spacetimes would require a way to decompose a function into a linear combination (possibly direct integral) of solutions of the wave equation with a specific value of the mass. This construction is possible on $\dS$ thanks to the Fourier transform, described in Section~\ref{sec_sols}.

The first intended application of this result is to formulate a scattering theory on~$\dS$, whether {\`a} la Haag--Ruelle or Araki--Haag, i.e., either constructing scattering states or calculating collision cross sections \cite{ArakiHaag67,DybalskiGerard14b,Haag58,Ruelle62}. However, both results on Minkowski spacetime are based on the spectral condition, the fact that the joint energy-momentum spectrum on the one-particle Hilbert space has an isolated point in its spectrum. A similar result does not exist on $\dS$. Actually, the plane waves written in the hyperbolic coordinate system (equations~\eqref{hyper-odd} and~\eqref{hyper-even}) are eigenfunctions of the total angular momentum operator and of one of its components (characterized by the quantum number~$m$), besides the mass operator. Moreover a one-particle Hilbert space is generated by the Fourier coefficients \eqref{F-hyper} for which $\rho \equiv \mu'$. However there is no element of the Lorentz algebra of which the plane wave is an eigenfunction and whose eigenvalue is related to the mass. The ``time evolution operator'' $\partial/\partial\beta$ does not represent an element of the Lorentz algebra. Therefore the wavepackets defined in this work are an important result in the pursuit of a scattering theory on a curved spacetime.

\appendix
\section{Appendix}\label{appendix}

We will calculate the parameter $d(\rho)$. This parameter arises from an intertwiner between homogeneous functions on $\fC$ given by the Fourier transform. The space of homogeneous functions of degree $\sigma=-\frac{(n-1)}{2}+\mri\rho$ on $\fC$ is generated by functions of the form $h(\zeta,\tau',{\bf \chi'}) = \zeta^\sigma\varpi(\tau',{\bf \chi'})$, $\varpi \in \cC^\infty\big({\pm}1 \times S^{n-1}\big)$. The space $\cC^\infty\big({\pm}1 \times S^{n-1}\big)$ is spanned by spherical harmonics and we will benefit from the fact that the Fourier transform of a spherical harmonic is a spherical harmonic~\cite{SteinWeiss71}. Hence we write
\begin{gather*}
\varpi(\tau',{\bf \chi'})= (\sign \tau' )^k Y_j({\bf \chi'}) ,
\end{gather*}
$k$ an $j$ being integer numbers, the degree of the spherical harmonics. $\varpi$ is even (odd) if $k+j$ is even (odd). Therefore
\begin{gather}
 Th(s,t',{\bf x'}) = \int_0^\infty\!\iint h(\zeta,\tau',{\bf \chi'})\mre^{\mri (\tau't'-{\bf \chi'}\cdot {\bf x'})s\zeta}\zeta^{n-2}\mrd \zeta \mrd\tau'\mrd^{n-1}\chi'\label{F-intertw} \\
\hphantom{Th(s,t',{\bf x'})}{} = \int_0^\infty \zeta^{\frac{n-3}{2}+\mri\rho}\left[\int_{S^{n-1}}\mre^{-\mri({\bf \chi'}\cdot {\bf x'})s\zeta}Y_j({\bf \chi'})\mrd^{n-1}\chi'\right]\left[\sum_{\tau'=\pm 1}\left(\sign \tau'\right)^k \mre^{\mri\tau't's\zeta}\right]\mrd\zeta . \nonumber
\end{gather}
Classical Fourier analysis gives
\begin{gather*}
\int_{S^{n-1}}\mre^{-\mri({\bf \chi'}\cdot {\bf x'})s\zeta}Y_j({\bf \chi'})\mrd^{n-1}\chi' = (2\pi)^{\frac{n}{2}}\mre^{-\mri\frac{j\pi}{2}}(s\zeta)^{-\frac{(n-2)}{2}}J_{\frac{n+2j-2}{2}}(s\zeta)Y_j({\bf x'})
\end{gather*}
and
\begin{gather*}
 \sum_{\tau'=\pm 1}\left(\sign \tau'\right)^k \mre^{\mri\tau't's\zeta} = \mre^{\mri t's\zeta}+(-1)^k\mre^{-\mri t's\zeta} = (\sign t')^k\mre^{\mri s\zeta} + \left(-\sign t'\right)^k\mre^{-\mri s\zeta} \\
\qquad{} = \sqrt{\frac{s\zeta\pi}{2}}\big\{\big[(\sign t')^k + (-\sign t' )^k\big]J_{-\frac{1}{2}}(s\zeta) +\mri\big[(\sign t')^k - (-\sign t' )^k\big]J_{\frac{1}{2}}(s\zeta)\big\} \\
\qquad{} =\begin{cases}
2\sqrt{\dfrac{s\zeta\pi}{2}}(\sign t')^k J_{-\frac{1}{2}}(s\zeta) & \textrm{if $k$ is even}, \vspace{1mm}\\
2\mri\sqrt{\dfrac{s\zeta\pi}{2}}(\sign t')^k J_{\frac{1}{2}}(s\zeta) & \textrm{if $k$ is odd}.
\end{cases}
\end{gather*}
Inserting these results in \eqref{F-intertw} one finds
\begin{gather*}
 Th(s,t',{\bf x'}) = \frac{1}{2}(2\pi)^{\frac{n+1}{2}}\mre^{-\mri\frac{j\pi}{2}}s^{-\frac{n-3}{2}}Y_j({\bf x'})
 \int_0^\infty \zeta^{\mri\rho}J_{\frac{n+2j-2}{2}}(s\zeta) \\
\hphantom{Th(s,t',{\bf x'}) =}{}
\times\big\{\big[(\sign t')^k + (-\sign t' )^k\big]J_{-\frac{1}{2}}(s\zeta) +\mri\big[(\sign t')^k - (-\sign t' )^k\big]J_{\frac{1}{2}}(s\zeta)\big\} \mrd\zeta \\
\hphantom{Th(s,t',{\bf x'}) }{} = \frac{(2\pi)^{\frac{n+1}{2}}}{2}\mre^{-\mri\frac{j\pi}{2}}s^{-\frac{n-1}{2}-\mri\rho}Y_j({\bf x'})
 \int_0^\infty y^{\mri\rho}J_{\frac{n+2j-2}{2}}(y) \\
\hphantom{Th(s,t',{\bf x'}) =}{}
 \times\big\{\big[(\sign t')^k + (-\sign t' )^k\big]J_{-\frac{1}{2}}(y) +\mri\big[(\sign t')^k - (-\sign t' )^k\big]J_{\frac{1}{2}}(y)\big\} \mrd y .
\end{gather*}
This integral may be evaluated by analytic continuation. If the integrand were $y^{-\lambda}J_\eta(y)J_\nu(y)$ and $\operatorname{Re}(\eta+\nu)+1 > \operatorname{Re}(\lambda)>0$, the integral would converge and the result would be an analytic function of $\lambda$ in the interval $\operatorname{Re}(\lambda) \in (0,\operatorname{Re}(\eta+\nu)+1)$ (see entry 24 in Table~19.2 of \cite{Bateman-tableII}). Thus by analytic continuation the result extends to $\operatorname{Re}(\lambda)=0$:
\begin{gather}
Th(s,t',{\bf x'}) = (2\pi)^{\frac{n+1}{2}}2^{\mri\rho}\mre^{-\mri\frac{j\pi}{2}}\times s^{-\frac{n-1}{2}-\mri\rho}\varpi(t',{\bf x'}) \nonumber \\
\hphantom{Th(s,t',{\bf x'}) =}{} \times \frac{\Upgamma(-\mri\rho)\Upgamma\left(\frac{n+2j \pm 1}{4}+\frac{\mri\rho}{2}\right)}{\Upgamma\left(\frac{-(n+2j) \pm 1}{4}+1-\frac{\mri\rho}{2}\right)\Upgamma\left(\frac{n+2j \pm 1}{4}-\frac{\mri\rho}{2}\right)\Upgamma\left(\frac{n+2j \mp 1}{4}-\frac{\mri\rho}{2}\right)}p(k) .\label{F-intertw-final}
\end{gather}
In the arguments of the Gamma functions, the upper sign must be chosen if $k$ is odd, and the lower sign if $k$ is even. Besides,
\begin{gather*}
p(k)=\begin{cases}
\mri & \textrm{if $k$ is odd}, \\
1 & \textrm{if $k$ is even}.
\end{cases}
\end{gather*}
We remark that the term $\frac{\Upgamma\left(\frac{n+2j \pm 1}{4}+\frac{\mri\rho}{2}\right)}{\Upgamma\left(\frac{n+2j \pm 1}{4}-\frac{\mri\rho}{2}\right)}$ is just a phase factor and denoting $z \coloneqq \frac{n+2j \mp 1}{4}-\frac{\mri\rho}{2}$, one finds that $\frac{-(n+2j) \pm 1}{4}+1-\frac{\mri\rho}{2} = 1-\overline{z}$. Therefore the other two Gamma functions in the denominator in \eqref{F-intertw-final} are
\begin{gather*}
\Upgamma(1-\overline{z})\Upgamma(z)=\overline{\Upgamma(1-z)}\Upgamma(z)=\Upgamma(1-z)\Upgamma(z)\mre^{-2\mri\Arg\left[\Upgamma(1-z)\right]}=\frac{\pi}{\sin\pi z}\mre^{-2\mri\Arg\left[\Upgamma(1-z)\right]} .
\end{gather*}
Collecting all phase factors in a term denoted as $p'(n,j,k,\rho)$, and comparing the result of \eqref{F-intertw-final} with \eqref{intertwiner}, one obtains
\begin{gather*}
 \Upgamma\left(\frac{n-1}{2}+\mri\rho\right)s^{-\frac{(n-1)}{2} - \mri \rho}\mre^{-\frac{\mri\pi}{2}\left(\frac{n-1}{2}+\mri\rho\right)} \\
\qquad\quad{} \times\iint|a|^{-\frac{n-1}{2}-\mri\rho}\left[\Theta(a)+\mre^{\mri\pi\left(\frac{n-1}{2}+\mri\rho\right)}\Theta(-a)\right] \varpi(\tau',{\bf \chi'})\mrd\tau'\mrd^{n-1}\chi' \\
\qquad{}=(2\pi)^{\frac{n+1}{2}}\frac{\sin\left[\pi\left(\frac{n+2j \mp 1}{4}-\frac{\mri\rho}{2}\right)\right]}{\pi}\Upgamma(-\mri\rho)p'(n,j,k,\rho)s^{-\frac{n-1}{2}-\mri\rho}\varpi(t',{\bf x'}).
\end{gather*}
Hence
\begin{gather}
\varpi(t',{\bf x'}) = (2\pi)^{-\frac{(n+1)}{2}}\frac{\Upgamma\left(\frac{n-1}{2}+\mri\rho\right)}{\Upgamma(-\mri\rho)}\frac{\pi}{\sin\left[\pi\left(\frac{n+2j \mp 1}{4}-\frac{\mri\rho}{2}\right)\right]}\mre^{\frac{\pi\rho}{2}}p''(n,j,k,\rho) \nonumber \\
\hphantom{\varpi(t',{\bf x'}) =}{} \times \iint|a|^{-\frac{n-1}{2}-\mri\rho}\left[\Theta(a)+\mre^{\mri\pi\left(\frac{n-1}{2}+\mri\rho\right)}\Theta(-a)\right] \varpi(\tau',{\bf \chi'})\mrd\tau'\mrd^{n-1}\chi' ,\label{intertw-d}
\end{gather}
where $p''(n,j,k,\rho)$ is just another phase factor.

Since the Fourier transform of a spherical harmonic is a spherical harmonic, comparing equations \eqref{intertw-unit} and \eqref{intertw-d} we conclude that the multiplicative parameter in the right-hand side of the latter is $d(\rho)$, whose absolute value is given in \eqref{d-parameter}.

\subsection*{Acknowledgements}
MB received financial support from the S\~ao Paulo Research Foundation (FAPESP) under grant $\#$2015/02975-4 during the preparation of this work. We would also like to thank the anonymous referees who made important comments on a previous version of this paper.

\pdfbookmark[1]{References}{ref}
\LastPageEnding

\end{document}